\documentstyle{article}
\begin{document}
\rightline{October 1992}
\rightline{Revised April, Aug 93, Jan 94}
\rightline{McGill/93-31}

\vskip 1.2cm
\begin{center}
\begin{large}
{\bf Explicit Electroweak symmetry breaking?}
\vskip 1.5cm
\end{large}
R. Foot$^{(a)}$ and Tran Anh Tuan$^{(b)}$
\end{center}
\vskip 1cm
\noindent
(a) Physics Department, McGill university, 3600 University
street, Montreal, H3A 2T8,
Canada\footnote{Present address} and Institute of Theoretical Physics,
Academy of Science of Vietnam, P. O. Box 429 Bo Ho, Hanoi 10000,
Vietnam.
\vskip 1cm
\noindent
(b) Graduate College in Physics,
National Centre for Scientific Research of Vietnam,
P.O. Box 429, Bo Ho, Hanoi 10000, Vietnam.

\vskip 2cm
\begin{center}
{\bf Abstract}
\end{center}
\vskip .6cm
\noindent
We hypothesise that all electroweak symmetry breaking terms
such as fermion masses and the W and Z gauge boson masses arise
radiatively from just one explicit symmetry breaking term in the Lagrangian.
Our hypothesis is motivated by the lack of experimental support
and the lack of predictivity of the standard symmetry breaking
scenario. We construct a simple model which illustrates our ideas.
We also discuss the possible importance of scale invariance.

\newpage
The standard model (SM) is a very successful model. It answers many
questions and explains many experiments. However, in some aspects it
seems to be unsatisfactory. In particular, the (SM) does not give
any understanding of the fermion mass hierarchy. It can
incorporate the empirical masses but there is no explanation of,
for e.g. why are $m_e, m_u << m_t$?
At present there are few fundamental problems in theoretical
physics in which experiment is ahead of theory. A quantitative
understanding of the fermion and gauge boson masses is, we believe,
one such fundamental problem.
A quantitative understanding of the fermion and gauge boson
masses is a difficult problem and if such an understanding could be
achieved it would undoubtly lead to a deeper insight into nature.

One of the obstacles in understanding the masses of the elementary
particles is that the masses are connected with the physics of
symmetry breaking (since the known fermion
and gauge boson mass terms do not respect the
electroweak gauge symmetry). The origin of electroweak symmetry
breaking is not known at present. The SM ascribes symmetry breaking to a vacuum
expectation value (VEV) of a fundamental scalar field.
This scalar couples to fermions and gauge bosons and the VEV
consequently gives tree-level masses to the fermions and bosons.
This sector of the SM does not have support yet from experiment.
The SM is simple but it is not compelling.

In this note we propose a completely different scenario
for the form of symmetry breaking. {\it We hypothesise that all gauge
symmetry breaking arises radiatively from just one explicit
symmetry breaking term in the Lagrangian.}[1]
We have no a priori
justification for this hypothesis. We propose to study it for
the following reasons. Firstly it is a simple hypothesis.
Secondly, it has not been previously studied (as far as we are aware) [2]
and most important, if the hypothesis is correct then it might be possible to
make some progress in understanding the fermion and gauge
boson mass hierarchy, since our hypothesis requires all fermion and gauge
boson masses to appear ultimately from just one source term
in the Lagrangian. Since we consider explicit symmetry breaking
rather than spontaneous symmetry breaking (as in the SM) our
model will be non-renormalizable. Our point of view is that the
divergences will be cancelled by some new and unknown physics,
possibly associated with the symmetry breaking source term.
Our approach is a bottom up approach. We do not propose `a theory of
everything' and we do not claim to know
all the answers.

What then is the one electroweak symmetry breaking term?
Perhaps the simplest
candidate for the symmetry breaking
term is for it to be a mass term (rather then some type of
interaction term).
This term could be either a gauge boson mass term or a fermion
mass term. If the symmetry breaking term is assumed
to be a gauge boson mass term then it seems to be impossible
to radiatively generate the fermion mass terms (due to
chiral symmetry). Thus we argue that if there is only one symmetry
breaking term in the Lagrangian, then a fermion mass
term is a good candidate.

Throughout this paper, we assume that the gauge symmetry of the
Lagrangian is the same as in the standard model, i.e.
$SU(3)_c \otimes SU(2)_L \otimes U(1)_Y$.
If symmetry breaking arises from just one tree-level
fermion mass term, then we expect
that this fermion mass, $M$, will satisfy
$$M \gg M_W, M_Z. \eqno (1)$$
Since the $W$ and $Z$ boson masses ($M_W, M_Z$) are by assumption assumed to
arise radiatively from the fermion mass $M$.
None of the known fermions
(or even the top quark) satisfy this condition so we expect that the
fermion mass term must be something exotic. Perhaps the simplest (or
at least most obvious) possibility is to
assume the existence of right-handed gauge singlet
Weyl fields and consider a large
Dirac tau neutrino mass term:
$$M_{\nu_{\tau D}} \gg M_W, M_Z.  \eqno (2)$$
We assume a see-saw type picture where there is a very large
Majorana mass for the right-handed neutrinos (so that the
physical left-handed neutrino mass is small enough to be
within the experimental bounds). It has already been shown [3],
that this case does lead to satisfactory one-loop induced
masses for the $W$ and $Z$ gauge bosons (which are proportional to
$M_{\nu_{\tau D}}$) and these masses automatically satisfy
the relation
$$ {M_W^2 \over M_Z^2 cos^2 \theta_w} \simeq 1, \eqno (3)$$
where the weak mixing angle, $\theta_w$ is defined by the ratio
of $SU(2)_L$ and $U(1)_Y$ gauge coupling constants i.e. $\tan \theta_w = g/g'$.
The one-loop diagrams are logarithmically
divergent however, so that a cutoff is required [4]. This,
unfortunately leads to a lack of predictivity, but we don't
consider it a problem for the low energy theory since
the cutoff is presumed to be associated with the physics
of the symmetry breaking term, or
some other type of unknown physics. Despite this lack of predictivity,
we think that in principle quantitative predictions can still be made
since the infinities can cancel if we consider
quantities such as mass ratios (for example).

Thus we argue that physics below $100$ GeV can be satisfactorily explained
by assuming that electroweak symmetry breaking is explicit (rather than
spontaneous). At high energies of the order of a TeV, the tree-level
elastic scattering cross section for longitudinal
$W$ and $Z$ bosons will violate the partial wave unitarity bound.
Thus, we expect the model to break down at the TeV scale.

We have not said anything yet about how the fermion masses can arise from
the symmetry breaking source term. In the model as it stands, there
is no mechanism for the fermions to gain
masses. Of course this is a crucial issue.
In this paper we focus
on illustrating our idea, and (like many other authors [2])
assume the existence of elementary scalars (which do not get any VEVs)
to communicate the symmetry breaking to the fermions.
Once we introduce elementary scalars, we also introduce arbitrariness
as in the case of the SM. This is clearly unsatisfactory
but we do not have many better ideas at the moment. Perhaps
a better alternative might be to consider the possibility
of a larger gauge symmetry than the SM, and to try and
generate the fermion masses with loops involving gauge bosons. This
possibility  is interesting, but not easy since we will need to extend
our hypothesis to include the additional symmetry breaking of the
larger gauge group. We hope to study this possibility sometime in the future
but
for the present, we assume only the SM gauge symmetry, and allow for the
possibility of scalars (with no VEVs) to generate the
fermion masses radiatively. This will allow us to illustrate our
idea.

Since the top quark mass is expected to be about the same
as the $W$ and $Z$ gauge boson masses (i.e. within a factor of 2),
we would also like to have this mass generated at one-loop. However
assuming the SM gauge symmetry, it is impossible to generate the top quark
mass from
$M_{\nu_{\tau D}}$ at one-loop using elementary scalars (with no
VEVs) to connect these two fermions together. To
see this, note that the Feynam diagram should have the form shown in
figures 1a or lb.
It is straightforward to show that such diagrams could
not exist.  To see this note that the first vertex in figure 1a requires a
$\bar t_L \chi \nu_{DR}$ interaction term
and the second vertex in figure 1a requires a
$\bar \nu_{DL}\chi^{\dagger}
t_R$ interaction term, but $\bar t_L \nu_{DR}$ does not transform like
$\bar t_R \nu_{DL}$ under the SM gauge symmetry
(and for the alternative possibility of the top quark being generated from
figure 1b, a similiar arguement holds).
One can of course construct a diagram with two
different $\chi$'s which are mixed
together by a symmetry breaking mass term. However in this
case the Lagrangian would contain two symmetry breaking
terms (and would consequently violate our hypothesis of only one
symmetry breaking term in the Lagrangian).

Thus we are led to consider a completely new type of fermion with a large mass
$M$ which is presumed to be the origin of symmetry breaking. If we add
a doublet of fermions and give one of them a large mass,
then the induced $W$ and $Z$ gauge boson masses do not automatically
satisfy eq. (3). However as already mentioned, if (for example)
the doublet of fermions contains a neutral fermion, and if one of the
Weyl components of the neutral fermion is a $SU(2)_L \otimes U(1)_Y$
singlet and has a large Majorana mass (which is just the see-saw
type picture), then the induced $W$ and $Z$ gauge boson masses do
automatically satisfy eq.(3). The Majorana mass term of the gauge singlet does
not violate our hypothesis since it is not a gauge symmetry breaking
term.

We start by allowing for the most general $SU(3)_c \otimes
SU(2)_L \otimes U(1)_Y$ transformation properties of the new fermion
$F_L, F_R$. Since $SU(3)_c$ is presumed unbroken,
$F_L, F_R$ must have identical $SU(3)_c$ transformation properties
(otherwise the $\bar F_L F_R$ mass term would break $SU(3)_c$).
For simplicity, we assume that they are color singlets. The simplest
possibility is then:
$$F_L \sim (1, M_1, y_L), \  F_{1R} \sim (1, 1, 0), \
F_{2R} \sim (1, M_1 -1, y_R). \eqno (4)$$
The one tree-level $SU(2)_L \otimes U(1)_Y$ symmetry breaking mass
term will be a
$M\bar F_{1L} F_{1R}$
 term in the Lagrangian (where $F_{1L}$ is one component of the multiplet
$F_L$)
i.e.
$${\cal L}_{SB} = M \bar F_{1L}F_{1R} + H.c. .\eqno (5)$$
As discussed earlier, we assume a large Majorana
mass term ${\cal M} \bar F_{1R} (F_{1R})^c$ (with ${\cal M} \gg M$)
for the gauge singlet. This leads to acceptable $W$ and $Z$ gauge boson
masses which are proportional to the fermion mass $M$ [3]. We also
require that the top quark gains mass at one-loop order. The required
Feynman diagram is shown in figure 2. In order to conect the
top quark with the fermion $F$ we assume the existence of a scalar $\chi$
(with no VEV). For the top quark to gain mass from
this diagram we must have the Lagrangian terms:
$${\cal L}_{\chi} = \lambda_1 \bar t_L F_{1R} \chi +
\lambda_2 \bar F_L t_R \chi^{\dagger} + H.c. .  \eqno (6)$$
The first term in the Lagrangian implies that
$$\chi \sim t_L \bar F_{1R} \sim (3, 2, 1/3). \eqno (7a)$$
The second term implies that
$$\chi \sim \bar F_L t_R \sim (3, M_1, 4/3 - y_L).  \eqno (7b)$$
So we see that $M_1 = 2$ and $y_L = 1$. Our notation is as follows:
$$F_L = \left(\begin{array}{c}
E\\
V
\end{array}
\right)_L \sim (1, 2, 1), \  F_{1R} = V_R \sim (1, 1, 0),\
F_{2R} = E_R \sim (1, 1, 2).  \eqno (8)$$
Thus assuming there exists a scalar $\chi$ with interactions
described by the Lagrangian terms ${\cal L}_{\chi}$ (eq.(6)), the
top quark (as well as the $W$ and $Z$ gauge bosons) gain
masses at one-loop order. All the other fermions remain massless
so we need to add more scalars. The exotic fermion $E$ must have
mass greater than about $45$ GeV, otherwise it would
have been detected in the LEP measurements
of the $Z$ boson width. Thus it is quite
heavy like the $W, Z$ gauge bosons and the top quark and we
would therefore expect its mass to be generated at the same
order in perturbation theory as the $W, Z$ gauge bosons and the top quark.
Thus the $E$ fermion should have its mass generated at one-loop, and this
requires the introduction of a new scalar $\phi$, with interactions:
$${\cal L}_{\phi 1} = \lambda_3 \bar F_L \phi V_R +
\lambda_4 \bar F_L \phi^c E_R + H.c. \eqno (9)$$
Note that, like the scalar $\chi$, the scalar $\phi$ is
assumed to have
no VEV. It is straightforward to show that this Lagrangian is
consistent with $SU(3)_c \otimes SU(2)_L \otimes U(1)_Y$ gauge
invariance provided
$$\phi \sim (1, 2, 1). \eqno (10)$$
 The Lagrangian terms ${\cal L}_{\phi 1}$ imply that the new
fermion $E$ gets its mass at one-loop, as shown in figure 3.
Given that $\phi \sim (1, 2, 1),$ it can also couple to the SM quarks
and leptons through the Lagrangian terms:
$${\cal L}_{\phi 2} = \lambda^1_{ij} \bar f_{iL} \phi e_{jR} +
\lambda^2_{ij} \bar f_{iL} \phi^c \nu_{jR} +
\lambda^3_{ij} \bar Q_{iL} \phi D_{jR}  $$
$$\  + \lambda^4_{ij} \bar Q_{ij} \phi^c U_{jR}
+ \lambda^5_i \bar f_{iL} \phi^c V_R +
\lambda^6_i \bar F_L \phi \nu_{iR} + H.c. \eqno (11)$$
with $i, j = 1, ... ,3$ representing the generations, i.e.
$$f_{1L} = \left(\begin{array}{c}
\nu_e\\
e
\end{array}\right)_L,\
f_{2L} = \left(\begin{array}{c}
\nu_{\mu}\\
\mu
\end{array}\right)_L,\
f_{3L} = \left(\begin{array}{c}
\nu_{\tau}\\
\tau
\end{array}\right)_L, \
  etc.
\eqno (12)$$
and our notation for the SM quarks should be self explanatory.
Note that we have also added gauge singlet right-handed neutrino
fields to the theory.
The Lagrangian terms ${\cal L}_{\phi 2}$ imply that
a Dirac mass term $M_{\nu_{\tau D}}$
is also generated at one-loop (and
one can check that no other fermion masses are generated at one-loop order).
This is shown in figure 4. Thus, with the scalars $\phi, \chi$ with
interactions
${\cal L}_\phi, {\cal L}_\chi$,
the $W, Z$, the top quark, the exotic $E$ fermion and a Dirac mass term
for $\nu_{\tau}$ are all generated at one-loop order in perturbation theory.
Each of these radiatively induced masses are directly proportional to
the symmetry breaking fermion
mass $M$. Note that we assume a large Majorana mass for the right-handed
neutrinos so that the physical neutrino masses are small (which is just
the well known see-saw
mechanism).
The simplest way that we have found to extend the
model so that all of the other quarks
and leptons get masses is to add another
copy of $\phi$ with interactions of the
same form as in eqs.(9) and (11). We denote the
two scalars as $\phi_1$ and $\phi_2$.
Assuming the most general Lagrangian for the interactions of the scalars
$\phi_1$ and $\phi_2$,
then in the quark sector,
at two-loops two charged $-1/3$ quarks will gain masses (which
we take to be the $b$ and $s$ quarks). At three-loops, the $c$ and
$u$ quarks will gain mass and the $d$ quark will get mass at four-loop order.
In the lepton sector, the $\tau$ and $\mu$ leptons get masses
at two-loops. At three-loops, the $\nu_{\mu}$ and $\nu_{e}$ get masses
and the electron gets mass at four-loop order in perturbation theory [5] [6].
Of course, the model as it stands can only partially explain the
fermion mass heirarchy and it certainly does not give
any quantitative mass predictions. Also note that the model is not complete
since it has triangle anomalies and there is also the requirement of a cutoff.
However the point of it is to illustrate our idea [7].

Finally, we would like to make some remarks concerning
scale invariance [8]. Perhaps an interesting possibility is that scale
invariance might also be explicitly broken at tree-level by just one term
in the Lagrangian. In particular, maybe both gauge and scale symmetry
breaking arises from just one source term [9].
It is not too difficult to modify our model so that this scenario is realized.
In our model, the only terms which break scale invariance at tree-level
are the masses for the scalars, the electroweak symmetry breaking mass
and the Majorana masses for the exotic $V_R$ as well as the $\nu_R$ fields.
The masses for the scalars will arise at one-loop if they are absent
in tree-level. The Majorana mass of $V_R$ maybe the source
of the majorana masses of the $\nu_R's$ (through one-loop diagrams
involving new scalars for example). The $V_R$ Majorana mass can be combined
with the electroweak symmetry breaking mass, so that the one
symmetry breaking term is then
$${\cal L}_{SB} = {\cal M} \bar V_R (\cos \phi (V_R)^c +
\sin \phi V_L) + H.c. \eqno (13)$$
We would also like to stress again that the symmetry breaking term
is not only the source of all the gauge and scale symmetry violating terms,
it is also the source of the non-renormalizability of
the theory. (This is unlike the situation in the SM where symmetry breaking
is spontaneous in origin and is renormalizable when terms with
(mass) dimension of less
than or equal to four are allowed at tree-level). We have two comments
on renormalizability.
Firstly, the model is not complete and some new physics must
exist somewhere which renders the divergent integrals finite.
Perhaps the symmetry breaking souce term should be viewed
as some effective term. Perhaps it is itself of dynamical origin
in which case there maybe no gauge or scale symmetry breaking
terms in the Lagrangian at all in the tree-level [10].
Secondly, it would be pleasing if renormalizability
followed from some principle. This is because, for example,
in nature there maybe some effective regulator which
renders the divergent integrals finite. In this case there is no reason
to exclude non-renormalizable terms from the tree-level Lagrangian.
However, if we start adding non-renomalizable terms to the
standard model, then some things would start to go wrong.
For example, the proton could decay and flavour changing
neutral currents could occur. What type of principle
could prevent non-renormalizable terms from appearing
in the Lagrangian of a gauge theory? Scale
invariance seems to be a good
candidate for such a principle [11].

In conclusion, we have considered the hypothesis that all electroweak symmetry
breaking is due to a single term
in the Lagrangian. All other symmetry breaking terms may arise
radiatively from this symmetry breaking term. From this assumption
we hope to make progress in understanding the fermion and gauge boson masses.
We constructed a simple model illustrating the idea.
However our
work is at present incomplete since the model contained triangle anomalies
and our Feynman diagrams were
logarithmically divergent. More importantly, we failed to make any mass
predictions, so we have not yet improved substantially upon the SM.
Nevertheless, we feel that in principle our idea could
be successful. For example, with some additional
assumptions, the Yukawa couplings of the scalars in our model
can be restricted so that quantitative predictions can be made.
Alternatively the use of scalars could be
abandoned and the gauge symmetry enlargened. Despite
the many shortcommings of our model, we think our idea is interesting.
It is important to explore new ideas in relation to the fermion mass
problem and we will continue to study this problem.
We also pointed out that the single symmetry
breaking term could be the only tree-level
source of scale symmetry breaking. We believe that scale invariance
may be an important concept since it is a symmetry
which forbids terms of (mass) dimension greater than four
in the Lagrangian (in the tree-level).

\vskip 1.8cm
\noindent
{\bf Acknowledgements}
\vskip 0.5cm
\noindent
R. Foot is grateful to Prof. Dao Vong Duc, Dr. Hoang Ngoc Long and
Dr. Nguyen Toan Thang and the staff of the Institute of theoretical
physics in Hanoi for their help and hospitality during a visit.
Tran Anh Tuan would like to thank Dr. Nguyen Suan Han and Dr. Hoang Ngoc
Long for encouragement.

\vskip 1cm
\noindent
{\bf Figure Captions:}
\vskip 0.4cm
\noindent
Figure 1a: Form of the Feynman diagram for radiatively
generating the top quark mass from the neutrino mass at one loop.
\vskip .4cm
\noindent
Figure 1b: Another possible form
for a Feynman diagram generating the top quark mass from the
neutrino mass at one loop.
\vskip .4cm
\noindent
Figure 2: Feynman diagram for the
radiative generation of the top quark mass from the
$V$ tree-level fermion mass term.
\vskip .4cm
\noindent
Figure 3: Feynman diagram for the radiative generation of the $E$ fermion
mass at one loop from the $V$ tree-level fermion mass term.
\vskip .4cm
\noindent
Figure 4: Feynman diagram for the radiative generation of a Dirac
tau neutrino mass term at one loop from the $V$ tree-level
fermion mass term.

\vskip 1cm
\noindent
{\bf References:}
\vskip .5cm
\noindent
[1]  Note however that there may also be some symmetry breaking due
to dynamical effects such as the formation of light quark
condensates which are believed to form in QCD. We assume that
such effects are negligible compared with the
symmetry breaking terms which are radiatively induced from
the tree-level source term.

\vskip .5cm
\noindent
[2] There has been a lot of studies discussing the possiblity of
generating some fermion masses from
others radiatively. However as far as we are aware, no-one has
considered our hypothesis before. For some earlier work on
radiative fermion mass generation see for example,
B. S. Balakrishna, Phys. Rev. Lett. 60, 1602 (1988);
X-G. He, R. R. Volkas, and D.-D. Wu, Phys. Rev. D41, 1630 (1990).
\vskip 0.5cm
\noindent
[3] R. Foot and S. Titard, Mod. Phys. Lett. A7 , 1991 (1992).
\vskip 0.5cm
\noindent
[4] Note that we assume that the cutoff is assumed to
preserve electroweak gauge symmetry.
This assumption is necessary, and it is consistent (at least
in spirt) with our hypothesis that there is only one Lagrangian
term which violates electroweak gauge symmetry.

\vskip .5cm
\noindent
[5] Note that the scalar masses can be arbitrarily large.
The model has no naturalness problem and no problem with
flavour changing neutral currents from scalar exchange (which
can be made arbitrarily small since the scalar masses can
be made arbitrarily large).
\vskip 0.5cm
\noindent
[6] The details of the mass generation of the quarks and the leptons
from the one-loop top quark and Dirac tau neutrino masses, will be presented
in more detail elsewhere [see R. Foot, H. N. Long and T. A. Tuan, to be
published.]. For this reason we have only sketched the details here.
\vskip 0.5cm
\noindent
[7] While our hypothesis is simple, there is of course no unique
implementation of it. In our example, we assumed that
the symmetry breaking term was a fermion mass term (eq.(5)).
Our model has 2 arbitrary mass scales (actually
5 when we include the Majorana masses for the right-handed neutrinos).
It might be interesting
to assume only one mass scale, say the Majorana mass scale $\cal M$.
We could then try to derive all the other masses in the theory
from this electroweak symmetric term by using a symmetry breaking
interaction term. The one mass scale could be very large
(e.g. the Planck mass) and the symmetry breaking term very small.
\vskip .5cm
\noindent
[8] Scale invariance can be defined mathematically as the
symmetry of the action under the scale transformations:
\begin{large}
$$x^{\mu} \rightarrow \lambda x^{\mu}, \ \  \{\phi, W^{\mu}, Z^{\mu}, A^{\mu}
\} \rightarrow \{\phi/\lambda, W^{\mu}/\lambda, Z^{\mu}/\lambda,
A^{\mu}/\lambda\},$$
$$   f \rightarrow f/\lambda^{{3 \over 2}}$$
\end{large}
where $f$ stands for the set of fermion fields.
\vskip .5cm
\noindent
[9] There may also be scale symmetry breaking due to dynamical effects
discussed in footnote [1].
\vskip .5cm
\noindent
[10] A. Zee [Phys. Rev. Lett. 44, 703 (1980)] has also speculated
that the ``theory of the world'' contains no dimensional parameter,
i.e. scale invariant. Zee's motivation for scale invariance seems
to be different from ours though.
\vskip .5cm
\noindent
[11] It is possible that non-renormalizable terms are present at the
tree-level. However, we expect that they must be heavily
suppressed due
to the observed stability of the proton. There are interesting exceptions
to this however. For example, there are some gauge models in which
the gauge symmetry is sufficient to forbid proton decay. See R. Foot,
Southampton University Preprint 1991 (unpublished) for details.

\end{document}